\documentstyle[amsmath,amssymb,aps,prl,multicol,graphicx,url]{revtex}

\newcommand{\ud}{\text{d}}
\newcommand{\ui}{\text{i}}
\newcommand{\ue}{\text{e}}

\newcommand{\erf}{{\rm erf}}

\newcommand{\Z}{{\mathbb{Z}}}

\newcommand{\kommentar}[1]{}

\begin{document}


\title{Level spacings and periodic orbits
  \begin{picture}(0,0) 
    \put(-359,30){\normalsize ULM-TP/00-8} 
    \put(-359,20){\normalsize December 2000, revised March 2001} 
  \end{picture}
}

\author{Stefan Keppeler\cite{emailkep}\\}
\address{
Abteilung Theoretische Physik, Universit\"at Ulm,
Albert-Einstein-Allee 11, D-89069 Ulm, Germany
}

\maketitle

\begin{abstract}
Starting from a semiclassical quantization condition based on the trace 
formula, we derive a periodic-orbit formula for the distribution of spacings
of eigenvalues with $k$ intermediate levels. Numerical tests verify the 
validity of this representation for the nearest-neighbor level spacing 
($k=0$). In a second part, we present an asymptotic evaluation 
for large spacings, where consistency with random matrix theory is achieved 
for large $k$. We also discuss the relation with the method of Bogomolny and 
Keating [Phys. Rev. Lett. {\bf 77} (1996) 1472] for two-point correlations.
\end{abstract}


\begin{multicols}{2}            

The statistical distribution of quantum energy levels is conjectured to 
reflect the chaoticity or integrability of the underlying classical dynamics
\cite{BerTab77b,BohGiaSch84}. For classically chaotic systems one expects
spectral distributions like those computed in random matrix theory (RMT)
\cite{BohGiaSch84}, whereas for classically integrable systems the quantum 
levels appear to follow the distribution for a Poisson process 
\cite{BerTab77b}. Attempts to
explain this correspondence are based on Gutzwiller's semiclassical 
trace formula \cite{Gut71} which bridges the gap between classical and quantum 
mechanics. The prime result in this direction is Berry's analysis of the
spectral rigidity \cite{Ber85}, based on the so-called diagonal 
approximation and the classical sum rule of Hannay and Ozorio de Almeida 
\cite{HanOzo84}. Starting from a semiclassical quantization condition 
\cite{AurSte92a,AurMatSieSte92}, Bogomolny and Keating \cite{BogKea96} were 
able to extract information on the two-point correlations going beyond the 
results of \cite{Ber85}. It is the aim of the present work to follow a 
similar path in order to obtain semiclassical information on the level 
spacing distributions.

The Gutzwiller trace formula can be expressed as a periodic-orbit sum for the 
integrated density of states $N(E):=\sum_n \Theta(E-E_n)$, 
\begin{equation}
\label{Gutz}
  N_T(E) \sim \bar{N}(E) 
  + \underset{|rT_\gamma| \leq T}{\sum \sum}^\prime 
  \frac{{\cal A}_{\gamma r}}{2\pi\ui r} \ue^{\frac{\ui}{\hbar} r S_{\gamma}(E)}
  \, ,
\end{equation}
in the semiclassical limit, i.e. for $\hbar \to 0$. $\bar{N}(E)$ denotes 
the mean part following from Weyl's law, whereas the fluctuations, 
$N_T^{\rm fl}(E)$, are given by a sum over all primitive periodic orbits 
$\gamma$ and their $r$-fold repetitions, 
the prime on the second sum indicating the exclusion of $r=0$. The (primitive)
action and period of a periodic orbit 
are denoted by $S_\gamma(E)$ and $T_\gamma(E)$, respectively.
The explicit form of the amplitudes ${\cal A}_{\gamma r}$ can be found in 
\cite{Gut71} for chaotic and in \cite{BerTab76} for integrable systems. 
We have given a version truncated at periods $|rT_\gamma| = T$, which will be 
needed in the following. Instead of directly using (\ref{Gutz}) to express 
spectral functions, we will use an approximate spectrum $E_n(T)$ obtained from
the condition \cite{AurSte92a,AurMatSieSte92} 
\begin{equation}
 N_T(E_n(T)) \stackrel{!}{=} n + \tfrac{1}{2} \, .
\end{equation}
Before investigating the statistical distribution we have to unfold the 
spectrum such that its mean density $\bar{d}(E) = \ud\bar{N}(E)/\ud E$
is rescaled to unity. To this end, investigating spectral 
correlations in an interval 
$I(E;\hbar):=[E-\hbar\omega,E+\hbar\omega]$, we introduce the unfolded 
energies $x_n(T):=E_n(T) \bar{d}$, $\bar{d}:=\bar{d}(E)$, see, e.g., 
\cite{BolKep99b} for details. Spacings of two unfolded energies with $k$ 
intermediate levels are given by 
$s_n(k;T)=x_{n+k+1}(T)-x_n(T)$ and integrated level spacing distributions 
are defined by 
\begin{equation}
\label{DefIvons}
  I(k,s;T) := \frac{1}{N_I} \sum_{E_n \in I(E;\hbar)} \Theta(s-s_n(k;T)) \, ,
\end{equation}
where $N_I$ denotes the number of eigenvalues contained in $I(E;\hbar)$.
The often used level spacing densities $P(k,s;T)$ are the derivative of 
$I(k,s;T)$ with respect to $s$. Since the condition 
$s \geq s_n(k;T)$ can be rewritten as 
$N_T(E_n(T)+s/\bar{d}\, ) - N_T(E_n(T)) \geq k$,
we can substitute the argument of the step-function.
Upon replacing the sum by an integral over $I(E;\hbar)$ with weight 
$d(E^\prime)$ which, in turn, we can asymptotically substitute by 
$d_T(E^\prime):=\ud N_T(E^\prime)/\ud E^\prime$ we obtain
\begin{eqnarray}        
\nonumber
  I(&&k,s;T) \sim \\ \label{Izwischen} && \bigg\langle
  \Theta \left( N_T \left( E^\prime + \frac{s}{\bar{d}} \right) 
               - N_T(E^\prime) - k - 1 \right) 
  \frac{d_T(E^\prime)}{\bar{d}} \bigg\rangle .
\end{eqnarray}  
The brackets denote the energy average 
$\langle \hdots \rangle := \frac{1}{2\hbar\omega} 
\int_{E-\hbar\omega}^{E+\hbar\omega} \hdots \ud E^\prime$,
due to which we may asymptotically replace 
$d_T(E^\prime)/\bar{d}$ by $1$. Since the semiclassical limit 
$\hbar \to 0$ now implies $\bar{d} \to \infty$, we have 
$s/\bar{d} \ll 1$, and expanding the mean part of $N_T$ about $E$ 
yields the periodic-orbit formula
\begin{equation}
\label{poIvons}
  I(k,s;T) \sim \left\langle 
  \Theta\left( s - k - 1 + N_T^{\rm fl} 
   \left( E^\prime + \frac{s}{\bar{d}} \right)
   - N_T^{\rm fl}(E^\prime) \right) \right\rangle 
\end{equation}
for the integrated level spacing distribution.
A similar expression can be obtained for the level spacing distribution 
$P(k,s;T)$ itself by taking the derivative of (\ref{poIvons}). Here we again 
neglect $d_T^{\rm fl}(E^\prime+s/\bar{d})/\bar{d}$ because of the 
energy average, i.e.
\begin{equation}
\label{poPvons}
  P(k,s;T) \sim \left\langle 
  \delta\left( s - k -1 + 
    N_T^{\rm fl} \left( E^\prime + \frac{s}{\bar{d}} \right)
    - N_T^{\rm fl}(E^\prime) \right) \right\rangle .
\end{equation}

\begin{figure}[H]
\begin{center}
\rotatebox{-90}{\includegraphics[width=6cm]{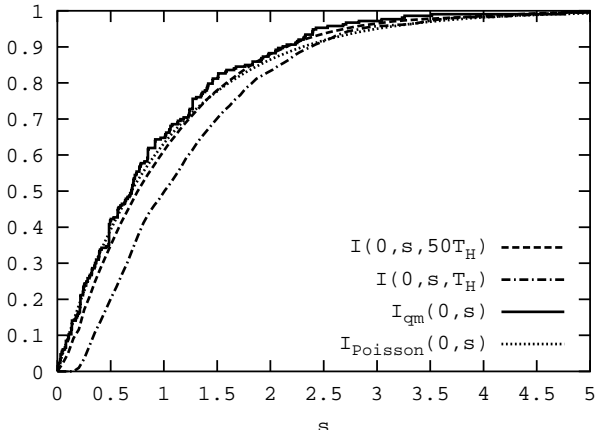}}
\begin{minipage}{8cm}
\caption{Integrated level spacing distribution from periodic orbits 
(\ref{poIvons}) for a rectangular billiard with cut-off time $50\, T_H$ 
(dashed), and $T_H$ (dashed-dotted). The quantum mechanical 
level spacing distribution (solid) and the graph for a 
Poisson process (dotted) are shown for comparison.}
\label{rechteck.plot}
\end{minipage}
\end{center}
\end{figure}

For the case $k=0$, i.e. for nearest-neighbor level spacings,
we will now test these formulae numerically for two toy models.
Our first example is a classically integrable system, namely a rectangular 
quantum billiard with aspect ratio $2/(1+\sqrt{5})$ and Neumann boundary 
conditions. For scaling systems the semiclassical limit $\hbar\to 0$ can 
be replaced by the high energy limit $E\to\infty$, with a suitable change 
in the energy average. For the plots we have taken into account energies 
in the interval $E\in[0,4000]$ (corresponding to 214 eigenvalues).
Figure \ref{rechteck.plot} shows the integrated level spacing distribution 
for this system  where for the cut-off time $T$ we have chosen 
$T=T_H:=2\pi\hbar\bar{d}$ (dashed-dotted line) and $T=50\, T_H$ (dashed line), 
respectively. When increasing 
the cut-off time $T$ we observe convergence towards the quantum mechanical 
result (solid line) which in turn in the limit $E\to\infty$ converges to
a Poissonian level spacing distribution (dotted line), cf. \cite{CheLeb91}
\vspace{-1ex}
\vspace{-1ex}

For a second test we have chosen the imaginary parts of the non-trivial 
zeros of the Riemann zeta function. These serve as a model for 
quantum chaos, since their spectral correlations are well described by 
the Gaussian unitary ensemble (GUE), see, e.g., \cite{BerKea99} and 
references therein. The density 
of the Riemann zeros is related to a sum over primes, see, e.g., \cite{Tit86},
in a similar way as the density of states of a quantum system is related to a 
sum over periodic orbits of the corresponding classical system (\ref{Gutz}). 
However, for chaotic systems as well as for the Riemann zeta function it has 
been observed that there might be an optimal cut-off for the periodic orbit 
sum, i.e. including arbitrarily long orbits doesn't necessarily improve the 
result. In the case of the Riemann zeta function this is known as the 
Riemann-Siegel cut-off \cite{Sie32} which corresponds to $T_H/2$ 
\cite{BerKea90}.

\begin{figure}[H]
\begin{center}
\rotatebox{-90}{\includegraphics[width=6cm]{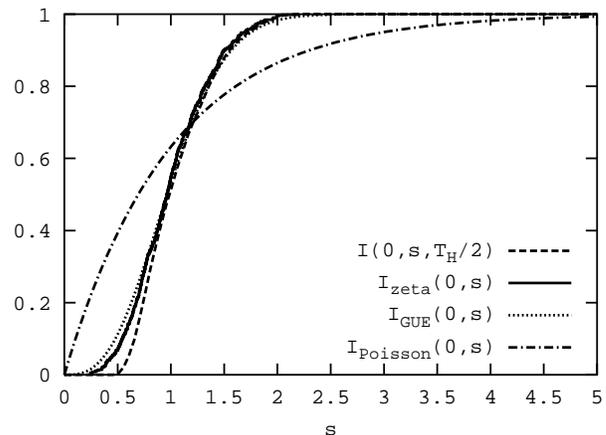}}
\begin{minipage}{8cm}
\caption{Integrated level spacing distribution for the zeros of the Riemann 
zeta function (solid) and the spacing distribution calculated from prime 
numbers (dashed), cf. (\ref{poIvons}), $T=T_H/2$. The spacing distribution of 
the GUE (dotted) and for a Poisson process (dashed-dotted) are shown for 
comparison.}
\label{zeta.plot}
\end{minipage}
\end{center}
\end{figure}

In figure \ref{zeta.plot} the integrated level spacing 
distribution of the first 649 Riemann zeros with imaginary 
parts between $0$ and $1000$ (solid line) is shown as well as the analogue 
of (\ref{poIvons}) (dashed line) with a truncation of 
the sum over primes at the Riemann-Siegel cut-off. 
We observe a good agreement of both curves at large values of $s$ were both
approach the GUE result (dotted line).
\vspace{-1ex}
\vspace{-2ex}
However, for values of $s$ below $1$ the periodic-orbit formula differs from 
both, the GUE and the exact curve. This is due to the fact that the periodic 
orbit sum (\ref{Gutz}), truncated at the order of the Heisenberg time,
cannot reproduce features below the scale of 
mean level spacing. Unlike in the integrable case, we cannot simply increase 
the cut-off since now the sum over primes behaves like an asymptotically 
divergent series, i.e. we observe 
best agreement when truncating the sum at the Riemann-Siegel 
cut-off. Analogously we will adopt the same point of view for periodic-orbit 
sums of classically chaotic systems by treating the respective 
trace formulae like asymptotically divergent series, choosing $T$ of the order 
of the Heisenberg time.

We remark that in numerical calculations it is necessary to use a smoothed
version of the periodic orbit sum (\ref{Gutz}). We have chosen to add a 
so-called remainder term as introduced by Aurich and Steiner \cite{AurSte92a}, 
which in the case of the Riemann zeta function is related to Guinand's 
formula \cite{Gui48}.

Now that we have checked the periodic orbit formula numerically, we attempt
to evaluate $P(k,s;T)$ asymptotically for generic quantum systems with 
chaotic classical limit. A well established conjecture on global eigenvalue 
statistics \cite{AurBolSte94} states that (after suitable normalization)
in the semiclassical limit the value distribution of $N_T^{\rm fl}$ is 
given by a Gaussian with 
zero mean. In analogy to \cite{BogKea96} we also assume Gaussian behavior
for the difference 
$N_T^{\rm fl}(E^\prime + s/\bar{d}\,) - N_T^{\rm fl}(E^\prime)$.
Although this assumption leads to the correct result in the case of 
two-point correlations \cite{BogKea96,Kep00}, we remark that it
corresponds to neglecting cross-correlations between $N_T^{\rm fl}$
at different arguments, which, especially for small $s$, will become important.
Thus, in the limit $s \to \infty$, $s/\bar{d} \to 0$,
the energy average in (\ref{poPvons}) can be performed approximately, yielding
\begin{equation}
\label{P_Gauss}
  P(k,s;T) \approx \frac{1}{\sqrt{2\pi\sigma^2}} 
  \exp\left( -\frac{(s-k-1)^2}{2\sigma^2} \right) \, ,
\end{equation}
where the ($s$-dependent) variance 
$\sigma^2 := \big\langle \big( N_T^{\rm fl}(E^\prime+s/\bar{d}) 
- N_T^{\rm fl}(E^\prime) \big)^2 \big\rangle$ of the Gaussian is still 
to be determined. This last step can be done by substituting the trace formula
(\ref{Gutz}) for $N_T^{\rm fl}$, employing an expansion in $s/\bar{d}$ 
(cf. \cite{Kep00}) and making use of the diagonal approximation 
\cite{HanOzo84,Ber85},
\begin{eqnarray}
  \sigma^2 && \sim \left( 
  \underset{|rT_\gamma|\leq T}{\sum\sum}^\prime
  \frac{{\cal A}_{\gamma r}}{2\pi\ui r} \, \ue^{\frac{\ui}{\hbar}r S_\gamma} 
  \left( \ue^{\frac{\ui}{\hbar}r T_\gamma \frac{s}{\bar{d}}} - 1 \right) 
  \right)^2 \\
  && \approx \underset{|rT_\gamma|\leq T}{\sum\sum}^\prime 
  \frac{g |{\cal A}_{\gamma r}|^2}{(2\pi r)^2} 
  \left( 2 - 2 \cos\left( \frac{s r T_\gamma}{\hbar\bar{d}} \right) \right) 
  \, ,
\end{eqnarray}
%
where $g$ denotes the generic multiplicity of orbits which share the 
same action, cf. \cite{Ber85,BolKep99b}. Using the classical sum rule 
\cite{HanOzo84} we obtain 
\begin{equation}        
  \sigma^2 \approx \frac{g}{\pi^2} \int_0^T 
  \frac{1 - \cos\left( \frac{sT^\prime}{\hbar\bar{d}} \right)}{T^\prime} 
  \, \ud T^\prime 
  \sim \frac{g}{\pi^2} \left( 
  \log\left( \frac{sT}{\hbar\bar{d}} \right) + \gamma \right) ,
\end{equation}  
$\gamma$ denoting Euler's constant. We remark that the 
sub-leading term of this expansion is not unique at this point, since it is 
influenced by corrections to the sum rule, which unfortunately are unknown, 
as well as by the choice of the cut-off time. Therefore, when now
setting $T = C T_H = C 2\pi\hbar\bar{d}$ we obtain 
\begin{equation} 
\label{sigma_result}
  \sigma^2 \approx \frac{g}{\pi^2} (\log s + \alpha) \, .
\end{equation}
Here $\alpha$ is kept as a free parameter, which, however, 
will not influence the large-$s$ asymptotics. 
Upon substituting (\ref{sigma_result}) into 
(\ref{P_Gauss}) we obtain a semiclassical formula for the level spacing 
distributions as $s\to\infty$,
\begin{equation}
\label{scPvons}
  P_{\rm sc}(k,s) \approx 
  \sqrt{\frac{\pi/(2g)}{\log s + \alpha}} \, 
  \exp\left( - \frac{\pi^2 (s-k-1)^2}{2g \, (\log s + \alpha)} \right) \, .
\end{equation}
Since (\ref{scPvons}) is mainly supported around $s \approx k+1$ we may for a
moment substitute $\log s$ by $\log(k+1)$ for large $k$. The formula obtained 
that way is consistent with a conjecture for $P(k,s)$ from RMT 
\cite{FreMelPan78}, see also \cite{BroFloFreMelPanWon81},
which can be adopted to generic quantum systems with 
chaotic classical limit \cite{AurSte90}.
However for small $k$ and in particular for nearest-neighbor level 
spacings (\ref{scPvons}) fails to reproduce the RMT results. This has to be 
understood in the following sense: The techniques used are essentially 
restricted to two-point correlations. Since the mean distance between 
$x_{n+k+1}$ and $x_n$ is given by $k+1$, the $P(k,s)$ can only be dominated by 
two-point correlations on scales where $s \approx k+1$. Thus an asymptotic
evaluation as $s\to\infty$ also implies large $k$.

We will now investigate the relation of our result (\ref{scPvons}) with the 
method of Bogomolny and Keating for the two-point correlation function 
$R_2(s)$, which can be represented as the sum 
\begin{equation}
\label{R2def}
  R_{2}(s) := \sum_{k=1}^\infty  P(k,s) - 1 \, .
\end{equation}
Substituting (\ref{scPvons}) for the level spacing distributions and applying 
the Poisson summation formula yields
\begin{eqnarray}\nonumber
  R_{2,{\rm sc}}(s) \approx \sum_{\nu \in \Z \setminus \{0\}} \frac{1}{2} 
    \exp(-2\pi^2\nu^2\sigma^2) \exp(2\pi\ui\nu s) \\ 
    \times \left[ 1 
    + \erf\left( \frac{s-1+2\pi\ui \nu \sigma^2}{\sqrt{2} \sigma} \right) 
    \right] \, ,
\end{eqnarray}
where $\sigma^2$ is given by (\ref{sigma_result}). The leading terms as 
$s\to\infty$ derive from $\nu=\pm1$ (as can be easily seen from the first 
exponential), i.e.
\begin{equation}
\label{R2sc}
  R_{2,{\rm sc}}(s) 
  \approx 2 \, \ue^{-2g\alpha} \, \frac{\cos(2\pi s)}{s^{2g}} \, .
\end{equation}
Comparing with the large-$s$ asymptotics of two-point correlation functions of
RMT we observe that we have obtained the leading oscillatory contribution, 
which can be expected from the method of Bogomolny and Keating 
\cite{BogKea96,Kep00}, but are missing the term which corresponds to the 
diagonal approximation of the spectral form factor, cf. 
\cite{Ber85,BogKea96,Haa00}. The reason for this is that between 
(\ref{Izwischen}) and (\ref{poPvons}) we have neglected the term 
$d_T^{\rm fl}(E^\prime)d_T^{\rm fl}(E^\prime+s/\bar{d})/\bar{d}^2$, 
which there was consistent, since 
it can be checked that doing so does not change the large-$s$ asymptotics 
of (\ref{scPvons}). However, in the sum (\ref{R2def}) the diagonal 
approximation of this term yields the missing term of $R_{2}(s)$. 

In this sense our result for the level spacing distributions (for large $k$)
is in leading order consistent with RMT two-point correlations. Thus, we can 
now compare (\ref{R2sc}) to the respective results from RMT, in order to fix 
$\alpha$. For systems without time-reversal symmetry ($g=1$) we have to 
compare with the GUE result, see, e.g. \cite{Meh91}, thus obtaining 
$\alpha = \log(2\pi)$. Analogously for time reversal invariant systems ($g=2$)
we obtain $\alpha = \log(\sqrt{2} \, \pi)$ by comparing with GOE two-point 
correlations.
Thus, with the substitution $\log s \approx \log(k+1)$ for large $k$ we 
have obtained Gaussians for the $P(k,s)$ with approximate variances
\[
  \sigma^2(k) \approx \frac{g}{\pi^2} (\log(k+1) + \alpha) \, .
\]

\kommentar{
In RMT one expects that $\sigma^2(k) \approx \Sigma^2(k+1) - 1/6$ for large 
$k$ \cite{FreMelPan78}, where 
$\Sigma^2(L):=\big\langle \big( N^{\rm fl}(E^\prime+L/\bar{d}) 
 - N^{\rm fl}(E^\prime) \big)^2 \big\rangle$ 
is the number variance. Using the large $L$ asymptotics for the number 
variances from RMT we can calculate the difference $\Sigma^2(k+1)-\sigma^2(k)$,
obtaining $0.160$ and $0.140$ for the GUE and the GOE, respectively. This has
just to be seen as a consistency check, since the constant $\alpha$ was not 
directly determined by semiclassics.
}

We briefly remark that the reasoning easily carries over to the third 
universality 
class, the Gaussian symplectic ensemble (GSE). To this end we have to consider 
trace formulae for particles with half integer spin \cite{BolKep98,BolKep99a}.
For time reversal invariant systems one has to take into account Kramers' 
degeneracy as in \cite{Kep00}. Again using the Gaussian ansatz 
(\ref{P_Gauss}) and methods of \cite{BolKep99b} for the calculation of 
$\sigma^2$ we obtain
\begin{equation}
\label{scPvons_spin}
  P_{\rm spin}(k,s) \approx 
  \sqrt{\frac{\pi}{\log s + \alpha}} \, 
  \exp\left( - \frac{\pi^2 (s-k-1)^2}{\log s + \alpha} \right) \, ,
\end{equation}
where the same procedure as above yields $\alpha = \log 8$. 

\kommentar{The difference 
$\Sigma^2(k+1)-\sigma^2(k) \approx 0.165$ is now in even better agreement with
the expected value $1/6$ than in the previous cases.}

Summarizing, we have derived periodic orbit formulae for the level spacing 
distributions $P(k,s)$ and $I(k,s)$. Numerical tests show that using purely 
classical input even the nearest-neighbor spacing distribution can be obtained 
for both, classically integrable and chaotic systems, where in 
the latter case we cannot go below the scale of mean level spacing. 
In a second part we have presented an asymptotic evaluation of the formulae 
for large spacings which yields good approximations, if we restrict ourselves 
to the case of large $k$, but does not apply to the nearest-neighbor 
level spacing.

It is a pleasure to acknowledge stimulating discussions with Dr. Jens Bolte,
Prof. Jon Keating, Roman Schubert and Prof. F. Steiner. This work was 
supported by Deutsche Forschungsgemeinschaft (DFG) under contract no. 
STE 241/10-1.

\kommentar{

}

\bibliographystyle{my_unsrt}                   
\bibliography{literatur}                       

\begin{thebibliography}{10}

\vspace{-1cm}
\bibitem[*]{emailkep}E-mail address: {\tt kep@physik.uni-ulm.de}

\bibitem{BerTab77b}
M.~V. Berry and M.~Tabor: {\em Level clustering in the regular spectrum\/},
  Proc. R. Soc. London Ser. A {\bf 356} (1977) ~375--394.

\bibitem{BohGiaSch84}
O.~{B}ohigas, M.-J. {G}iannoni and C.~{S}chmit: {\em {C}haracterization of
  chaotic quantum spectra and universality of level fluctuation laws\/}, Phys.
  Rev. Lett. {\bf 52} (1984) ~1--4.

\bibitem{Gut71}
M.~C. {G}utzwiller: {\em {P}eriodic {O}rbits and {C}lassical {Q}uantization
  {C}onditions\/}, J. Math. Phys. {\bf 12} (1971) ~343--358.

\bibitem{Ber85}
M.~V. {B}erry: {\em {S}emiclassical theory of spectral rigidity\/}, Proc. R.
  Soc. London Ser. A {\bf 400} (1985) ~229--251.

\bibitem{HanOzo84}
J.~H. Hannay and A.~M. {Ozorio de Almeida}: {\em Periodic orbits and a
  correlation function for the semiclassical density of states\/}, J. Phys. A
  {\bf 17} (1984) ~3429--3440.

\bibitem{AurSte92a}
R.~{A}urich and F.~{S}teiner: {\em {S}taircase functions, spectral rigidity,
  and a rule for quantizing chaos\/}, Phys. Rev. A {\bf 45} (1992) ~583--592.

\bibitem{AurMatSieSte92}
R.~{A}urich, C.~{M}atthies, M.~{S}ieber and F.~{S}teiner: {\em {N}ovel {R}ule
  for {Q}uantizing {C}haos\/}, Phys. Rev. Lett. {\bf 68} (1992) ~1629--1632.

\bibitem{BogKea96}
E.~B. Bogomolny and J.~P. Keating: {\em {G}utzwiller's {T}race {F}ormula and
  {S}pectral {S}tatistics: {B}eyond the {D}iagonal {A}pproximation\/}, Phys.
  Rev. Lett. {\bf 77} (1996) ~1472--1475.

\bibitem{BerTab76}
M.~V. {B}erry and M.~{T}abor: {\em {C}losed orbits and the regular bound
  spectrum\/}, Proc. R. Soc. London Ser. A {\bf 349} (1976) ~101--123.

\bibitem{BolKep99b}
J.~{B}olte and S.~{K}eppeler: {\em {S}emiclassical form factor for chaotic
  systems with spin 1/2\/}, J. Phys. A {\bf 32} (1999) ~8863--8880.

\bibitem{CheLeb91}
Z.~Cheng and J.~L. Lebowitz: {\em Statistics of energy levels in integrable
  quantum systems\/}, Phys. Rev. A {\bf 44} (1991) ~R3399--R3402.

\bibitem{BerKea99}
M.~V. Berry and J.~P. Keating: {\em The {R}iemann zeros and eigenvalue
  asymptotics\/}, SIAM Rev. {\bf 41} (1999) ~236--266.

\bibitem{Tit86}
E.~C. Titchmarsh: {\em The theory of the {R}iemann zeta-function\/}, The
  Clarendon Press Oxford University Press, New York, 2nd edn.,  (1986), Edited
  and with a preface by D. R. Heath-Brown.

\bibitem{Sie32}
C.~L. Siegel: {\em {\"U}ber {R}iemanns {N}achla\ss\ zur analytischen
  {Z}ahlentheorie\/}, {Q}uellen und {S}tudien zur {G}eschichte der {M}ath.,
  {A}str. und {P}hysik {\bf 2} (1932) ~45--80.

\bibitem{BerKea90}
M.~V. Berry and J.~P. Keating: {\em A rule for quantizing chaos?\/}, J. Phys. A
  {\bf 23} (1990) ~4839--4849.

\bibitem{Gui48}
A.~P. Guinand: {\em A summation formula in the theory of prime numbers\/},
  Proc. London Math. Soc. (2) {\bf 50} (1948) ~107--119.

\bibitem{AurBolSte94}
R.~{A}urich, J.~{B}olte and F.~{S}teiner: {\em {U}niversal {S}ignatures of
  {Q}uantum {C}haos\/}, Phys. Rev. Lett. {\bf 73} (1994) ~1356--1359.

\bibitem{Kep00}
S.~Keppeler: {\em Two-point correlations of the {G}aussian symplectic ensemble
  from periodic orbits\/}, J. Phys. A {\bf 33} (2000) ~L503--L507.

\bibitem{FreMelPan78}
J.~B. French, P.~A. Mello and A.~Pandey: {\em Statistical Properties of
  Many-Particle Spectra. II. Two-Point Correlations and Fluctuations\/}, Ann.
  Phys. (NY) {\bf 113} (1978) ~277--293.

\bibitem{BroFloFreMelPanWon81}
T.~A. Brody, J.~Flores, J.~B. French, P.~A. Mello, A.~Pandey and S.~S.~M. Wong:
  {\em Random-matrix physics: spectrum and strength fluctuations\/}, Rev.
  Modern Phys. {\bf 53} (1981) ~385--479.

\bibitem{AurSte90}
R.~{A}urich and F.~{S}teiner: {\em {E}nergy-{L}evel {S}tatistics of the
  {H}adamard-{G}utzwiller {E}nsemble\/}, Physica D {\bf 43} (1990) ~155--180.

\bibitem{Haa00}
F.~{H}aake: {\em {Q}uantum {S}ignatures of {C}haos\/}, {S}pringer-{V}erlag,
  {B}erlin {H}eidelberg,  (2000).

\bibitem{Meh91}
M.~L. {M}ehta: {\em {R}andom {M}atrices\/}, Academic Press, San Diego, 2nd
  edn.,  (1991).

\bibitem{BolKep98}
J.~{B}olte and S.~{K}eppeler: {\em {S}emiclassical Time Evolution and Trace
  Formula for Relativistic Spin-1/2 Particles\/}, Phys. Rev. Lett. {\bf 81}
  (1998) ~1987--1991.

\bibitem{BolKep99a}
J.~{B}olte and S.~{K}eppeler: {\em A semiclassical approach to the {D}irac
  equation\/}, Ann. Phys. (NY) {\bf 274} (1999) ~125--162.

\end{thebibliography}


\begin{references}
\vspace{-1.5cm}
\bibitem[*]{emailkep}E-mail address: {\tt kep@physik.uni-ulm.de}
\end{references}

\end{multicols}
\end{document}